\documentclass[
showpacs,
floatfix,
aps,
prl,
twocolumn,
superscriptaddress,
amssymb,
]{revtex4-1}

\usepackage{amssymb}
\usepackage{latexsym}
\usepackage[dvips]{graphicx}
\usepackage{amsmath}

\usepackage{graphicx}
\usepackage{dcolumn}
\usepackage{amsfonts}
\usepackage{bm}
\usepackage{epsfig}

\newcommand{\be}{\begin{equation}}
\newcommand{\ee}{\end{equation}}
\newcommand{\bea}{\begin{eqnarray}}
\newcommand{\eea}{\end{eqnarray}}

\def\edc{\epsilon_j}

\def\oc{\omega_{\mbox{\scriptsize {c}}}}

\def\rc{R_{\mbox{\scriptsize {c}}}}

\def\tpi{\tau_{\rm sh}}

\def\tq{\tau_{\mbox{\scriptsize {q}}}}
\def\ttr{\tau}

\def\tsh{\tau_{\rm sh}}

\newcommand{\req}[1]{Eq.\,(\ref{#1})}

\newcommand{\rfig}[1]{Fig.\,\ref{#1}}

\newcommand{\rref}[1]{Ref.\,\onlinecite{#1}}

\begin{document}
\title{Hall field-induced resistance oscillations in MgZnO/ZnO heterostructures}
\author{Q.~Shi}
\affiliation{School of Physics and Astronomy, University of Minnesota, Minneapolis, Minnesota 55455, USA}
\author{M.~A.~Zudov}
\email[Corresponding author: ]{zudov@physics.umn.edu}
\affiliation{School of Physics and Astronomy, University of Minnesota, Minneapolis, Minnesota 55455, USA}
\author{J.~Falson}
\affiliation{Max-Planck-Institute for Solid State Research, Heisenbergstrasse 1, D-70569 Stuttgart, Germany}
\affiliation{RIKEN Center for Emergent Matter Science (CEMS), Wako 351-0198, Japan}
\author{Y.~Kozuka}
\affiliation{Department of Applied Physics and Quantum-Phase Electronics Center (QPEC), The University of Tokyo, Tokyo 113-8656, Japan}
\author{A.~Tsukazaki}
\affiliation{Institute for Materials Research, Tohoku University, Sendai 980-8577, Japan}
\affiliation{PRESTO, Japan Science and Technology Agency (JST), Tokyo 102-0075, Japan}
\author{M.~Kawasaki}
\affiliation{Department of Applied Physics and Quantum-Phase Electronics Center (QPEC), The University of Tokyo, Tokyo 113-8656, Japan}
\affiliation{RIKEN Center for Emergent Matter Science (CEMS), Wako 351-0198, Japan}
\author{K.~von~Klitzing}
\affiliation{Max-Planck-Institute for Solid State Research, Heisenbergstrasse 1, D-70569 Stuttgart, Germany}
\author{J.~Smet}
\affiliation{Max-Planck-Institute for Solid State Research, Heisenbergstrasse 1, D-70569 Stuttgart, Germany}

\begin{abstract}
We report on nonlinear magnetotransport in a two-dimensional electron gas hosted in a MgZnO/ZnO heterostructure. 
Upon application of a direct current, we observe pronounced Hall field-induced resistance oscillations (HIRO) which are well known from experiments on high-mobility GaAs/AlGaAs quantum wells. 
The unique sensitivity of HIRO to the short-range component of the disorder potential allows us to unambiguously establish that the mobility of our MgZnO/ZnO heterostructure is limited by impurities residing within or near the 2D channel. 
Demonstration that HIRO can be realized in a system with a much lower mobility, much higher density, and much larger effective mass than in previously studied systems, highlights remarkable universality of the phenomenon and its great promise to be used in studies of a wide variety of emerging 2D materials.

\end{abstract}
\maketitle

Assessing the nature and magnitude of the disorder present in high-quality two-dimensional electron systems (2DESs) presents a challenging yet necessary experimental task. 
When cooled to a low temperature $T$ and exposed to a magnetic field $B$, these systems are widely known to divulge a rich array of quantum phenomena which display both quantitative and qualitative dependencies on the underlying disorder. 
Unfortunately, standard magnetotransport measurements in isolation give limited insight into this disorder as not all carrier scattering events are reflected equally or separably in the measured resistance \citep{dmitriev:2012}. 
However, a better glimpse of key characteristics of the disorder potential may often be gained from non-equilibrium transport phenomena, such as microwave-induced resistance oscillations (MIRO) \citep{zudov:2001b,ye:2001} and Hall field-induced resistance oscillations (HIRO) \citep{yang:2002,bykov:2005c,zhang:2007a}. 
Both phenomena exploit the commensurability of the energy spacing between the centers of disorder-broadened Landau levels and either the photon energy of the incident radiation (MIRO) or the Hall voltage drop across the cyclotron orbit under applied direct current (HIRO).
Importantly, the amplitudes of these oscillations and their $B$-dependencies contain information on specific scattering types \citep{dmitriev:2005,vavilov:2007,khodas:2008,dmitriev:2009b}.

MIRO, appearing when a 2DES is exposed to microwave radiation of frequency $\omega = 2\pi f$ and a weak $B$-field, are controlled by $ \omega/\oc$, where $\oc = eB/m^\star$ is the cyclotron frequency of an electron with the effective mass $m^\star$. 
To date, MIRO have been observed in GaAs/AlGaAs \citep{zudov:2001b,note:zrs}, Ge/SiGe \citep{zudov:2014}, and MgZnO/ZnO \citep{karcher:2016} heterostructures. 
Usually \citep{note:altTheories}, MIRO are explained in terms of the displacement \citep{ryzhii:1970,durst:2003,lei:2003,vavilov:2004} and the inelastic \citep{dorozhkin:2003,dmitriev:2005} contributions. 
The former originates from the radiation-induced modification of scattering off impurities and carries important information about correlation properties of the disorder potential.
The latter accounts for the radiation-induced changes to the distribution function and is controlled by the ratio of the transport and electron-electron scattering rates. 
As a result, the relative importance of these contributions depends on many parameters and their unequivocal disentanglement is a challenging experimental feat that remains to be accomplished \citep{note:sp}.
Moreover, the inelastic contribution can completely mask the displacement contribution in high-density and low-mobility 2DESs, such as the MgZnO/ZnO samples used in a recent MIRO study \citep{karcher:2016}, making it virtually impossible to extract correlation properties of the disorder potential from microwave photoresistance.

Another prominent non-equilibrium phenomenon, HIRO, emerges when a sufficiently strong direct current $I$ is sent through a 2DES placed in a varying $B$-field \citep{yang:2002,bykov:2005c,zhang:2007a}. 
HIRO appear as $1/B$-periodic oscillations in the differential resistance $r$ and originate solely from the displacement contribution from the electron backscattering off short-range (``sharp'') disorder \citep{vavilov:2007,lei:2007}.
Theoretically, the oscillatory correction to $r$ is given by \citep{vavilov:2007}
\be 
\frac {\delta r}{R_0} \approx \frac{16}{\pi}\frac{\tau}{\tau_{\pi}}\lambda^2 \cos 2\pi \edc \,,~~ \pi\edc \gg 1\,,
\label{eq.hiro} 
\ee
where $R_0$ is the low-temperature, linear-response resistance at $B=0$, $\tau$ is the disorder-limited transport scattering time, $\tau_{\pi}$ is the backscattering time, $\lambda = e^{-\pi/\oc\tq}$, 
$\tq$ is the quantum lifetime, $\edc = 2eE\rc/\hbar\oc$, $E = \rho_H I/w$ is the Hall field ($w$ is the sample width, $\rho_H$ is the Hall resistivity), and $\rc$ is the cyclotron radius.

For the exploration of disorder characteristics HIRO are unique because, in contrast to MIRO which are also sensitive to the radiation intensity and inelastic relaxation, the HIRO amplitude is controlled solely by the backscattering rate $\tpi^{-1}$ and the quantum scattering rate $\tq^{-1}$, entering $\lambda$.
The former characterizes only the short-range (``sharp'') component of the disorder potential while the latter also accounts for scattering off the long-range (``smooth'') disorder component. 
To date, HIRO have been studied in modulation-doped systems based on either GaAs/AlGaAs \citep{yang:2002,bykov:2005c,zhang:2007a,zhang:2007c,zhang:2008,hatke:2009c,hatke:2010a,hatke:2011a,hatke:2012d} or, more recently, on Ge/SiGe \citep{shi:2014b} heterostructures. 
Both of the studied systems are characterized by very high mobility ($\mu \sim 10^6 - 10^7$ cm$^2$/Vs), low effective mass ($m^\star \approx 0.06 -0.09 m_0$), and moderate carrier density (typically $n_e \sim 10^{11}$ cm$^{-2}$).

\begin{figure}[t]
\includegraphics{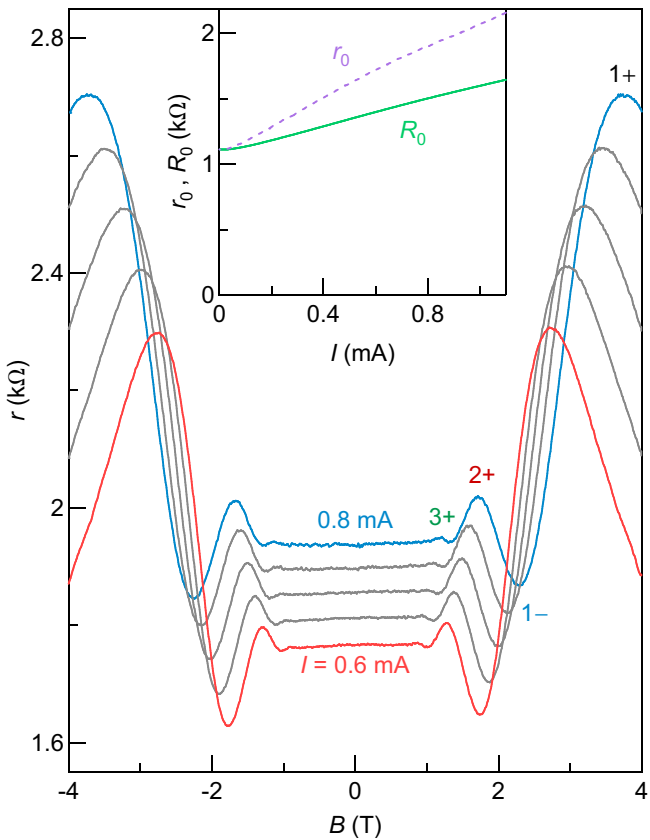}
\vspace{-0.05 in}
\caption{(Color online)
Differential magnetoresistance $r(B)$ measured at different $I$ from 0.6 to 0.8 mA, in steps of 0.05 mA.
The traces are \emph{not} vertically offset.
Inset: Zero-field differential resistance $r_0$ (dashed line) and total resistance $R_0$ (solid line) as a function of $I$.
At $I >$ 0.05 mA, $R_0(I)$ is well described by $R_0(I) = R_0(0) + a I$, with $R_0(0) \approx 1.1$ k$\Omega$, $a \approx 0.52$ k$\Omega$/mA.
}
\vspace{-0.15 in}
\label{fig1}
\end{figure}

In this Rapid Communication we report on the first observation and study of HIRO in a 2DES hosted in a MgZnO/ZnO heterostructure, a system which is distinct from both GaAs/AlGaAs and Ge/SiGe.
More specifically, the 2DES in our MgZnO/ZnO heterostructures is characterized by a much lower mobility ($\mu \sim 10^4$ cm$^2$/Vs), a much larger effective mass ($m^\star \approx 0.3m_0$), and much higher carrier density ($n_e \sim 10^{12}$ cm$^{-2}$). 
Despite these differences, our experiments reveal well-developed HIRO demonstrating that neither low effective mass nor high mobility of previously studied systems are essential for HIRO detection.
This finding highlights remarkable universality of the effect and its great promise to be realized in other 2D systems allowing their characterization. 
Taking our MgZnO/ZnO system as an example, we demonstrate that the analysis of the $B$-dependence of the HIRO amplitude allows us to unambiguously establish that its mobility is limited by short-range disorder originating from impurities at or near the interface.

Our sample was fabricated from a Mg$_{0.15}$Zn$_{0.85}$O/ZnO heterostructure grown using liquid ozone-based molecular beam epitaxy \cite{falson:2011,falson:2016,note:000}. 
A Hall bar mesa with a width of about 0.15 mm and a distance between voltage probes of about 0.8 mm was defined by scratching the wafer with a tungsten needle. 
Electrical contacts were made by soldered indium.
At low temperature, our 2DES has a carrier density $n_e \approx 2.0 \times 10^{12}$ cm$^{-2}$ and a mobility $\mu \approx 2.3 \times 10^4$ cm$^2$/Vs.
The differential magnetoresistance $r$ was measured at a fixed coolant temperature $T \approx 1.35$ K using a standard four-terminal lock-in detection scheme at various direct currents up to 1 mA.

In the main panel of \rfig{fig1}(a) we present $r$ as a function of $B$ recorded at different $I$ from 0.6 to 0.8 mA, in steps of 0.05 mA. Despite a much lower mobility of our sample compared to previously studied systems in which HIRO we observed to date, the data readily reveal well-developed HIRO which persist up to the third order (cf. $1+,2+,3+$) and expand to higher $B$ with increasing $I$.
It is important to note that, in contrast to GaAs/AlGaAs and Ge/SiGe in which HIRO typically occur at $B \sim 0.1$ T, HIRO in our ZnO sample can be extended to fields beyond 5 T (at $I \ge 1$ mA).
We notice that in view of high carrier density, HIRO are still observed in the regime of high filling factors (at $B = 5$ T we estimate $\nu \approx 17$).

\begin{figure}[t]
\includegraphics{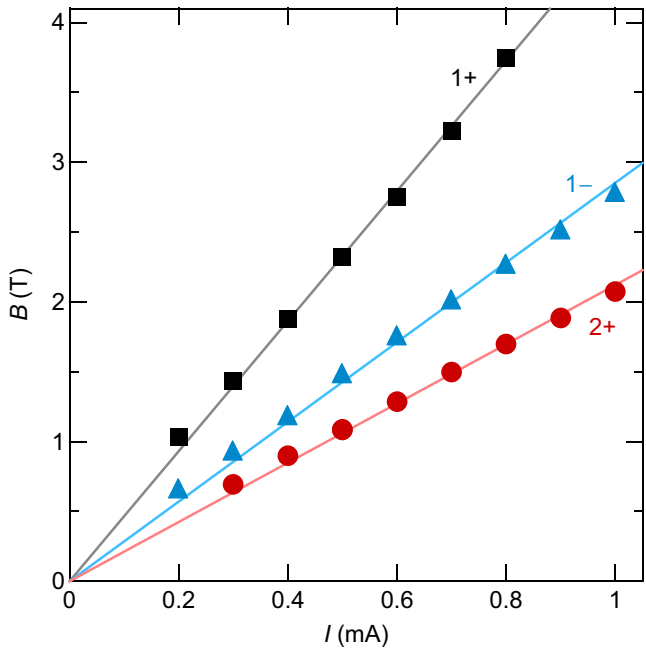}
\vspace{-0.05 in}
\caption{(Color online)
Magnetic fields at the oscillation extrema, $B_1^+$ (circles), $B_1^-$ (squares), and $B_2^+$ (triangles) as a function of $I$.
Lines are fits to the data.
}
\vspace{-0.15 in}
\label{fig2}
\end{figure}

According to \req{eq.hiro}, positions of the $n$-th HIRO maximum ($B_n^+$) and minimum ($B_n^-$) can be described by
\be
B_n^+ \approx \sqrt{\frac{8\pi}{n_e}}\frac {m^\star}{e^2} \frac j n ~ {\rm and} ~ B_n^- \approx \sqrt{\frac{8\pi}{n_e}}\frac {m^\star}{e^2} \frac j {n+1/2}\,,
\label{eq.bn}
\ee
where $j = I/w$ is the current density.
In \rfig{fig2} we present $B_1^+$ (circles), $B_1^-$ (squares), and $B_2^+$ (triangles) as a function of $I$ and observe linear dependencies, in accord with \req{eq.bn}.
For high carrier densities ($n_e \gtrsim 10^{12}$ cm$^{-2}$), as in our sample, both Shubnikov-de Haas oscillations \cite{falson:2015b} and MIRO \citep{karcher:2016} yield the effective mass close to the band mass $m^\star \approx 0.3 m_0$. 
Using this value and the obtained linear dependencies, we estimate the effective width of our Hall bar to be $w\approx 0.09$ mm, in reasonable agreement with the estimated scratch-defined width \citep{note:6}.
\begin{figure}[t]
\includegraphics{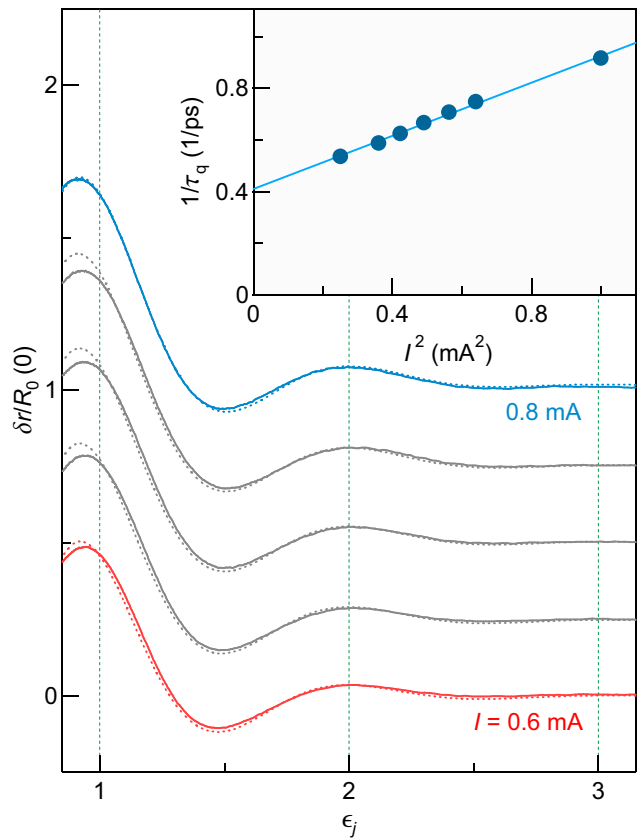}
\vspace{-0.05 in}
\caption{(Color online)
Measured $\delta r/R_0(0)$ (solid lines) and fits to the data with \req{eq.bessel} and $\ttr/\tpi = 1$ (dotted lines) as a function of $\edc$ for $I$ = 0.6 to 0.8 mA, in steps of 0.05 mA.
Traces are offset by 0.25 for clarity.
Inset: Quantum scattering rate $\tau_q^{-1}$ as a function of $I^2$.
Solid line is drawn at $\tq^{-1} =  \tq^{-1}(0) + bI^2$, with $\tq(0) \approx 2.4$ ps, $b \approx 0.51$ ps$^{-1}$/mA$^2$.
}
\vspace{-0.15 in}
\label{fig3}
\end{figure}

Further examination of \rfig{fig1} reveals that the zero-field differential resistance $r_0$ increases with $I$ and that, concurrently, the HIRO onset moves to higher $B$.
These observations suggest that both the transport and the quantum lifetime decrease with $I$.
The likely cause of these findings is the current-induced increase of the electron temperature due to Joule heating.
In the inset of \rfig{fig1} we present the differential resistance $r_0$ (dashed line) and the total resistance $R_0$ \citep{note:2} (solid line) as a function of $I$.
The total resistance can be described well by $R_0 = R_0(0) + a I$, $R_0(0) \approx 1.1$ k$\Omega$, $a \approx 0.52$ k$\Omega$/mA, indicating that the transport scattering time $\tau$ decreases by about one third at the maximum current $I = 1$ mA. 

The quantum lifetime and its dependence on $I$ can also be extracted. 
In view of a rather small number of observed oscillations, we opt for a direct fit of the experimental traces with the theoretical expression, as opposed to the conventional Dingle analysis \citep{zhang:2007a,hatke:2009c,hatke:2011a,hatke:2012d}.
More specifically, we examine the experimentally obtained relative oscillatory correction to the differential resistance $\delta r/R_0(0) \equiv (r-r_0)/R_0(0)$.
This quantity is shown in the main panel of \rfig{fig3} (solid lines) as a function of $\edc$ for the same currents as in \rfig{fig1}.
The traces, offset for clarity by 0.25, exhibit the same period and the same phase, showing that our analysis is self-consistent. 
The first-order minima and the second-order maxima occur close to $\edc = 1.5$ and $\edc = 2$, respectively, while the fundamental HIRO maxima appear at $\edc$ somewhat lower than unity.
This deviation is anticipated given the approximate nature of \req{eq.hiro} applicable only in the limit of $\pi\edc \gg 1$. 
We therefore chose to fit our experimental data with the full HIRO expression \citep{vavilov:2007},
\be
\frac {\delta r}{R_0} = -\frac{2\tau}{\tsh} \lambda^2 \left ( \zeta  \left [ J_0^2(\zeta) \right ]'' \right )'\,,
\label{eq.bessel}
\ee
where $J_0$ is the Bessel function and prime denotes the derivative over $\zeta = \pi \edc$.

Since $\lambda^2 = \exp(-2\pi\edc/\omega_j\tq)$, the only fitting parameters are $\tq^{-1}$ and $\ttr/\tpi$. 
We first notice that the fits consistently yield $\ttr/\tpi \approx 1.0$ (with an accuracy of a few percent) for all $I$, unambiguously signaling the prevalence of sharp-disorder (large-angle) scattering in our MgZnO/ZnO heterostructure. 
This finding is consistent with the close values of $\tq$ and $\ttr$ obtained in \rref{falson:2015b}, which suggested a less significant role of smooth-disorder (small-angle) scattering than in traditional, remotely-doped 2DES.
We therefore conclude that the electron mobility in our system is limited by impurities at or near the interface.
The most likely source of this scattering is the alloy disorder in Mg$_x$Zn$_{1-x}$O.

Since the ratio $\ttr/\tpi$ is expected to be the same for all $I$, we chose to fix its value at $\ttr/\tpi = 1$ and then perform single-parameter fits to obtain quantum scattering rate.
The obtained fits to the data are included in \rfig{fig3} as dotted lines, demonstrating excellent overlap with the experimental traces \citep{note:app}. 
The quantum scattering rate $\tq^{-1}$, extracted from the fits, is shown in the inset of \rfig{fig3} as a function of $I^2$. 
As illustrated by solid line, it can be described well by $\tq^{-1} =  \tq^{-1}(0) + bI^2$, with $\tq(0) \approx 2.4$ ps and $b \approx 0.51$ ps$^{-1}$/mA$^2$.
This observed increase with $I$ is likely caused by enhanced electron-electron scattering \citep{hatke:2009a,hatke:2009c} as the electron temperature increases.
The obtained value of $\tq(0)$ is in good agreement with the values extracted from Shubnikov-de Haas oscillations \cite{falson:2015b} and MIRO \citep{karcher:2016} measured in similar samples.

In summary, we have observed and investigated Hall field-induced resistance oscillations in a MgZnO/ZnO heterostructure, a recently developed 2DES.
By exploiting the direct sensitivity of HIRO to the sharp component of the disorder potential, we identified large-angle scattering off impurities within or near the interface as the dominant source of scattering \citep{note:last}. 
Since our MgZnO/ZnO sample has very different parameters than conventional high-mobility modulation-doped 2DESs in which HIRO have been observed so far, our findings demonstrate that HIRO is a powerful experimental tool to assess disorder characteristics across a wide variety of 2D systems.
In particular, we establish that neither low effective mass nor high mobility are prerequisites for reliable HIRO detection.

\begin{acknowledgements}
We thank I. A. Dmitriev for discussions and comments on the manuscript, P. Herlinger for assistance with the microscope, and A. Zudova for assistance with data acquisition.
This work was supported, in part, by the US Department of Energy, Office of Basic Energy Sciences, under Grant No. DE-SC002567 (University of Minnesota) and by Grant-in-Aids for Scientific Research (S) No. 24226002 from MEXT, Japan (University of Tokyo).
\end{acknowledgements}


\begin{thebibliography}{50}
\bibitem[{not({\natexlab{d}})}]{note:000}
\bibinfo{note}{We notice that the carrier-inducing mechanism in our sample
  relies not on a remote doping, as in earlier studied GaAs and Ge-based
  systems, but on a difference in polarization fields of a capping
  Mg$_x$Zn$_{1-x}$O and the underlying ZnO epilayer. As a consequence, the
  ZnO/MgZnO heterostructure is void of intentionally ionized impurities.}
\bibitem[{not({\natexlab{g}})}]{note:app}
\bibinfo{note}{This is quite remarkable since \req{eq.bessel} (as well as
  \req{eq.hiro}) was derived assuming $\oc\ttr \gg 1$ and $\oc\tq \lesssim 1$.
  In typical GaAs/AlGaAs samples $\ttr \gg \tq$ and both of these conditions
  can be simultaneously satisfied. Situation is markedly different in our ZnO
  sample where $\tq(0) \sim \ttr(0) \approx 4$ ps and the above conditions are
  only marginally met.}
\end{thebibliography}
\end{document}